\documentclass{article}
\usepackage{spconf}
\usepackage{amsmath,graphicx}
\usepackage{verbatim}

\title{Example-based super-resolution for point-cloud video}
\name{Diogo C. Garcia, Tiago A. Fonseca and Ricardo L. de Queiroz\thanks{Work partially supported by CNPq under grant 308150/2014-7. This paper was submitted to ICIP-2018 and its copyright may be transferred to IEEE.}}
\address{Universidade de Brasilia\\
Brasilia, Brasil\\
\normalsize \emph{\{diogo,tiago\}@image.unb.br and queiroz@ieee.org}}

\begin{document}
\maketitle
\begin{abstract}


We propose a mixed-resolution point-cloud representation and an example-based super-resolution framework, from which several processing tools can be derived, such as compression, denoising and error concealment.
By inferring the high-frequency content of low-resolution frames based on the similarities between adjacent full-resolution frames, the proposed framework achieves an average 1.18 dB gain over low-pass versions of the point-cloud, for a projection-based distortion metric~\cite{proposal:queiroz_torlig_fonseca:jpeg_2018,article:queiroz_chou:tip2016}.
\end{abstract}

\vspace{-2 ex}

\section{Introduction}
\label{sec:intro}


Recent demand for AR/VR applications have accelerated the interest in electronic systems to capture, process and render 3D signals such as point clouds~\cite{article:holoportation, man2_origin}. 
Nonetheless, there are no established standards regarding the capture, representation, compression and quality assessment of point clouds (PC). 
This lack of standards attracted attention to this research field and motivated a sequence of recent advances in processing 3D signals.

These signals can be captured using a set of RGBD cameras and represented as a voxelized point cloud. A voxelized PC consists in a set of points $(x, y, z)$ constrained to lie on a regular 3D grid~\cite{msf_seqs, man2}. 
Each point can be considered as the address of a volumetric element, or voxel, which is said to be occupied or unoccupied and, for each occupied position, the surface color (RGB) is recorded.
Instead of using a dense volumetric signal with all possible RGB samples for each frame, the point-cloud can be represented by a list of occupied voxels (geometry) and its color attributes, from now on referred as point-cloud.

\begin{figure}[h]
\centering
\begin{minipage}[t]{\columnwidth}
\centering
\centerline{\includegraphics[width=0.8\columnwidth]{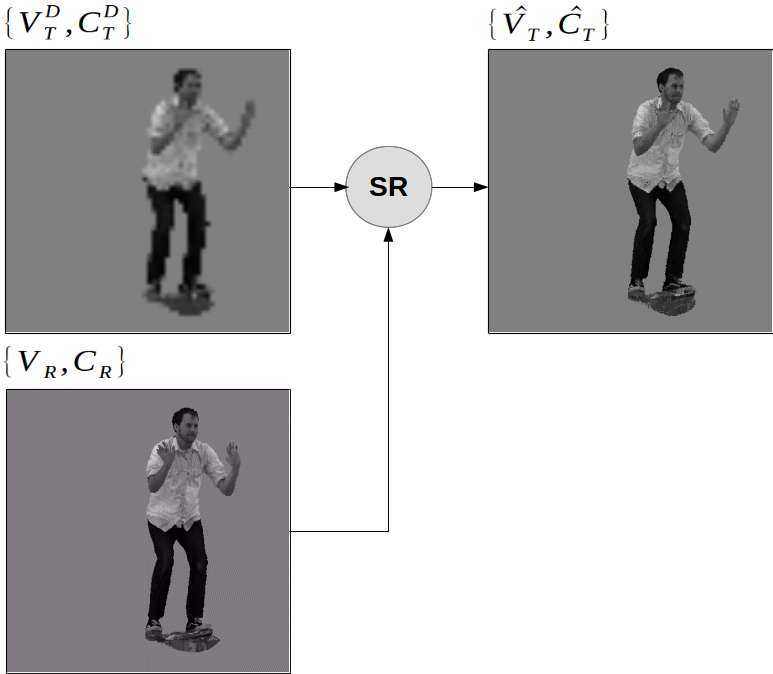}}
\end{minipage}
\vspace{-5 ex}
\caption{\small The super-resolution framework (SR) outputs a super-resolved frame $\{\hat{\mathbf{V}}_T, \hat{\mathbf{C}}_T\}$ by exploring the high-frequency content similarities between a down-sampled point-cloud $\{\mathbf{V}^D_T, \mathbf{C}^D_T\}$ and a time-adjacent frame $\{\mathbf{V}_R, \mathbf{C}_R\}$.}
\label{fig:sr_in_an_example}
\end{figure}

A very efficient geometry representation can be obtained using the octree method, where the 3D space is recursively divided into fixed-size cubes, or octants, allowing for data compression, fast searching and spatial scalability~\cite{man2_origin, man2, octree}.
Mixed-resolution scenarios (interleaved low- and full-resolution frames) naturally emerge from this representation: a low-resolution error-protected base layer, for instance, can be enhanced from a previously decoded full-resolution frame~\cite{article:hung_et_all:tcsvt2012, article:hung_garcia_queiroz:SPL2014}.
The super-resolution technique was already explored when processing 3D signals, with works trying to increase data
resolution in depth maps~\cite{article:mesquita_campos:sibgrapi2015, article:Ganihar:icvgip2014, article:diskin_asari:spie2012}. 
They usually explore geometric properties to improve the level of detail of a depth map. 
The contribution brought by this work is to infer high-frequency content (detail) of a point cloud by exploring the similarities between time-adjacent frames of already voxelized point-cloud signals as illustrated in Fig.~\ref{fig:sr_in_an_example}.
A concise signal model is derived in Section~\ref{sec:prop} and drives an example-based super-resolution framework, which is able to enhance the point-cloud level of detail as shown in Section~\ref{sec:results}. The conclusions are presented in Section~\ref{sec:conc}.

\vspace{-2 ex}

\vspace{-4 ex}
\section{Proposed method}
\label{sec:prop}

Traditional super-resolution (SR) \cite{article:hung_et_all:tcsvt2012} techniques generate high-resolution images from either multiple low-resolution images (multi-image SR) or databases of low- and high-resolution image pairs (example-based SR). The proposed method borrows ideas from the latter in order to increase the resolution and the details of a low-resolution point-cloud. Instead of relying on a database of low- and high-resolution point-cloud pairs, the proposed method extracts high-resolution information from an adjacent frame in a mixed-resolution point-cloud video sequence.

In this scenario, each point-cloud frame is composed of a pair of lists $\{\mathbf{V}, \mathbf{C}\}$, representing the XYZ positions of the occupied voxels and their corresponding color channels, respectively. Each down-sampled target frame $\{\mathbf{V}^D_T, \mathbf{C}^D_T\}$ is preceded by a full-resolution reference frame $\{\mathbf{V}_R, \mathbf{C}_R\}$. The proposed method generates the super-resolved version of the target frame, $\{\hat{\mathbf{V}}_T, \hat{\mathbf{C}}_T\}$, by adding the high-resolution information from the reference frame to the downsampled target frame.

In order to extract high-resolution information from $\{\mathbf{V}_R, \mathbf{C}_R\}$, the algorithm may estimate the motion of the voxels from one frame to another. However, only a downsampled version of the target frame is available. By downsampling the reference frame, $\{\mathbf{V}^D_R, \mathbf{C}^D_R\}$, motion can be estimated in a lower resolution of the point-clouds. However, the resolution of the motion estimation is also lowered. For example, if nearest-neighbour downsampling by a factor $D_F$ of $2$ is employed, the motion estimation is not be able to find any 1-voxel motion at full resolution, regardless of the motion direction.

\begin{figure}[htb]
\includegraphics[width=\columnwidth]{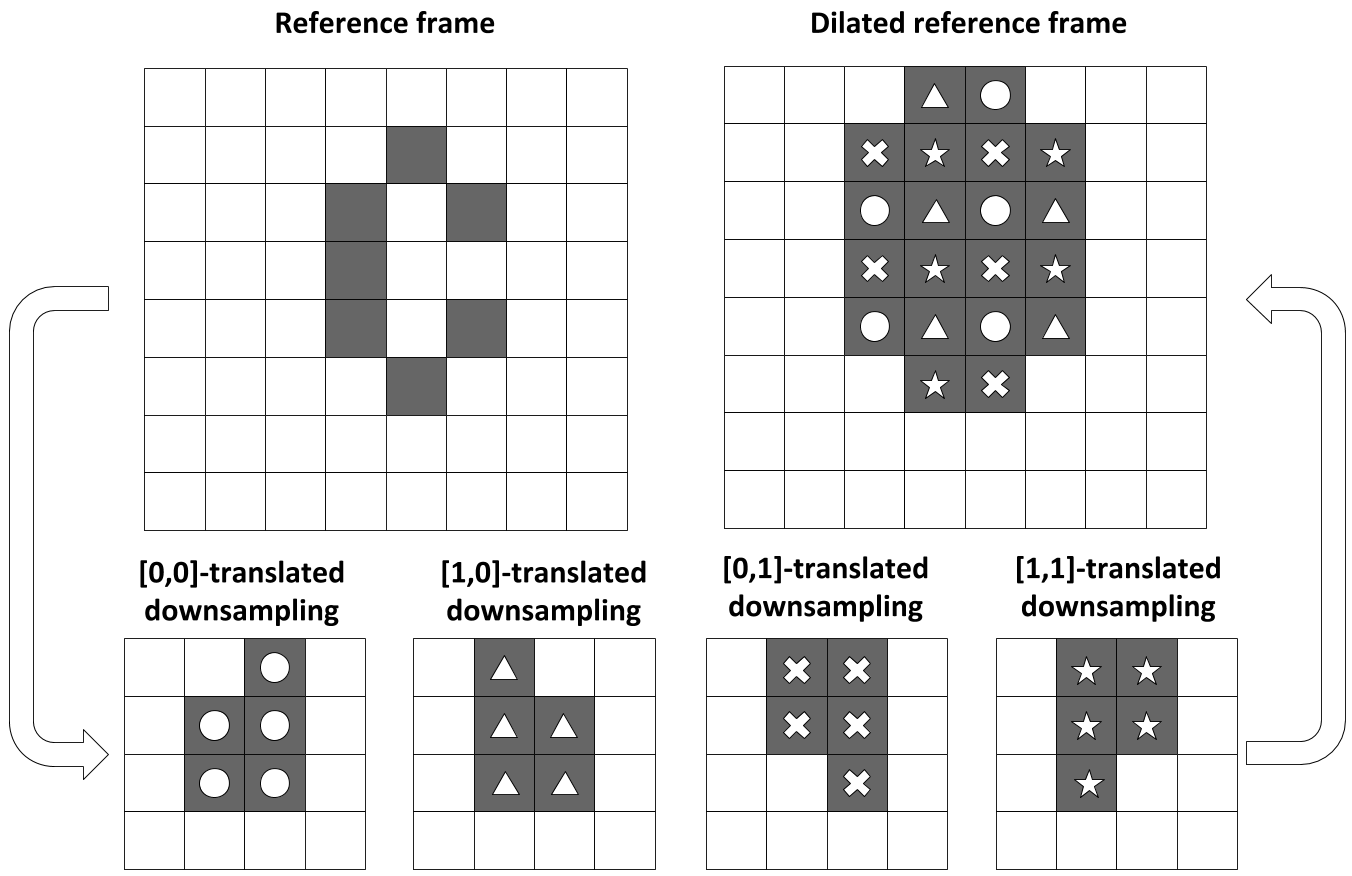}
\caption{\small Downsampling of a 2D symbol by a factor of $2$, considering 1-pixel translation in the XY directions, and posterior grouping, which can be seen as a dilation of the original symbol with a $2 \times 2$ square. The circles, squares, multiplication signs and stars illustrate from which downsampled version each position in the dilated version came from.}
\label{fig:phase_down}
\end{figure}

A better estimation of the point-cloud's motion can be achieved by generating several downsampled versions of the reference frame considering incremental motion in all directions. For example, for $D_F=2$, $8$ downsampled versions can be generated by considering XYZ translations by $[0,0,0]$, $[0,0,1]$, $[0,1,0]$, $[0,1,1]$, $[1,0,0]$, $[1,0,1]$, $[1,1,0]$ and $[1,1,1]$. At full resolution, this is equivalent to dilating the reference point-cloud by a $2 \times 2 \times 2$ cube, rendering $\mathbf{V}^{DL}_R$. Figure \ref{fig:phase_down} illustrates this concept in a 2D scenario, where only XY translations are considered, and dilation is performed with a $2 \times 2$ square.

\begin{figure}[htb]
\includegraphics[width=\columnwidth]{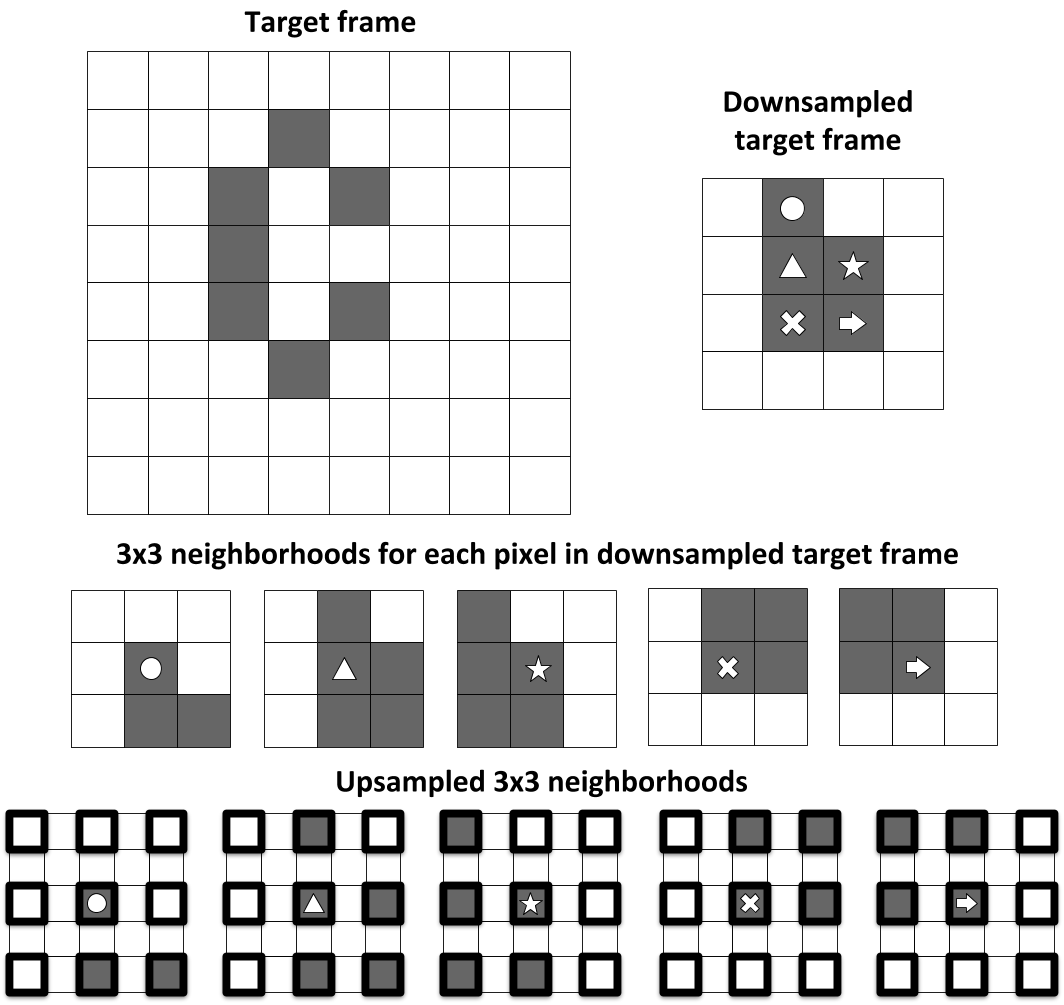}
\caption{\small $3 \times 3$ neighborhood around the downsampled version of a 2D symbol by a factor of $2$, and the upsampled version of these neighborhoods. Note that only every other position is considered in the upsampled version, as indicated by the thicker lines around the considered pixels.}
\label{fig:tgt_neigh}
\end{figure}

Super-resolution of the target frame requires that $D_F \times D_F \times D_F$ voxels should be estimated for each voxel in $\mathbf{V}^D_T$. In order to do that, motion estimation is performed between $\mathbf{V}^D_T$ and $\mathbf{V}^{DL}_R$ on a voxel basis, using the $N \times N \times N$ neighborhood around each voxel as support. Each neighborhood $\mathbf{n}^D_T$ and $\mathbf{n}^{DL}_R$ is defined as a binary string, indicating which voxels are occupied and which are not. Since $\mathbf{V}^D_T$ and $\mathbf{V}^{DL}_R$ have different resolutions, the neighborhoods in these frames are obtained by skipping $D_F-1$ positions around each of their voxels, which acts as the motion estimation between $\mathbf{V}^D_T$ and each of the downsampled versions of $\mathbf{V}_R$. Figure \ref{fig:tgt_neigh} illustrates this concept in a 2D scenario for a $3 \times 3$ neighborhood.

In the motion-estimation process, an $L \times L \times L$ search window around each voxel is chosen by the user, creating a tradeoff between speed and super-resolution accuracy. The cost function $C(i,j)$ between the $i$-th voxel in $\mathbf{V}^D_T$ and the $j$-th voxel in $\mathbf{V}^{DL}_R$ is defined as:

\begin{equation}
C(i,j) = \frac{H(i,j) + wD(i,j)}{w+1},
\end{equation}

\noindent where $H(i,j)$ is the hamming distance between the neighborhoods $\mathbf{n}^D_T(i)$ and $\mathbf{n}^{DL}_R(j)$, $D(i,j)$ is the Euclidean distance between $\mathbf{V}^D_T(i)$ and $\mathbf{V}^{DL}_R(j)$, and $w$ is the inverse of the Euclidean distance between the centers of mass of $\mathbf{V}^D_T$ and $\mathbf{V}^{DL}_R$. A small Euclidean distance between these centers of mass indicates small motion between the reference and target frames, increasing the value of $w$, i.e. $C(i,j)$ favors smaller motion vectors.

\vspace{-2 ex}

\subsection{Performance Metrics}
\label{subsec:performance_metric}

This work uses two measures to evaluate the achieved signal enhancements: a geometric quality metric~\cite{article:gpsnr_vetro:icip2017}, referred as GPSNR, and a distortion metric referred as projection peak signal to noise ratio (PPSNR)~\cite{article:queiroz_chou:tip2017,proposal:queiroz_torlig_fonseca:jpeg_2018}. 
GPSNR uses point-to-plane distances between point-clouds to calculate the geometric fidelity between them. 
The following steps are performed to evaluate the PPSNR:
\begin{enumerate}	
\item Each target frame is expected to be represented in three versions: original $\{\mathbf{V}_{T}, \mathbf{C}_{T}\}$, low-pass $\{\mathbf{V}^{DL}_{T},\mathbf{C}^{DL}_{T}\}$, and super-resolved $\{\hat{\mathbf{V}}_{T},\hat{\mathbf{C}}_{T}\}$ point-clouds. 
For each version, project ($\pi_{i}(\mathbf{\{V,C\}}) = f(x, y)$) the 3D signal on the faces of its surrounding cube to get 6 2D signals, $\Pi(\mathbf{\{V,C\}}) = \bigcup_{i=1}^{6} \pi_{i}(\mathbf{\{V,C\}})$, for each point-cloud version. Each projection $\pi_{i}(\mathbf{\{V,C\}})$ outputs a 512$\times$512 YUV~\cite{book:richardson:2004} color image.
Those represent the scene views from an orthographic projection on each cube face.
\item For each cube face, evaluate the luma PSNR between the projections of the original $\pi_{i}(\{\mathbf{V}_{T},\mathbf{C}_{T}\})$ and the super-resolved $\pi_{i}(\{\hat{\mathbf{V}}_{T},\hat{\mathbf{C}}_{T}\})$ signals and the projections of the original and the upsampled $\pi_{i}(\{\mathbf{V}^{DL}_{T},\mathbf{C}^{DL}_{T}\})$ signals. 
The PSNR between projections will provide 6 PSNRs values for each signal pair.
\item Average the 6 PSNRs to get two quality assessments: one evaluating the super-resolved and another evaluating the low-pass version. 
Then, subtracts the first value and the second to get the SR quality enhancement. 
\item To get a PNSR value which represents the SR derived improvements, take the average of the PSNR differences along the sequence frames.
\end{enumerate}

\vspace{-3 ex}

\section{Experimental results}
\label{sec:results}

Tests for the proposed method were carried out with seven point-cloud sequences: \textit{Andrew}, \textit{David}, \textit{Loot}, \textit{Man}, \textit{Phil}, \textit{Ricardo} and \textit{Sarah} \cite{msf_seqs, 8iseqs:2017, man2}.
The test set is made of five upper-body scenes of subjects captured at 30 fps and recorded at a spatial resolution of 512$\times$512$\times$512 voxels or a 9-level resolution. \textit{Man} and \textit{Loot} are full-body scenes recorded at 9- and 10-level resolution, respectively.
The average PPSNR was calculated according to Sec.~\ref{subsec:performance_metric}.

Table \ref{tab:res1} summarizes the PPSNR performance of the proposed SR method for the test set. 
The comparison metric is the difference between the average PPSNR of the SR method and that evaluated for the low-pass version for each sequence. 
Table~\ref{tab:res2} shows the average PPSNR and GPSNR performance gains.
\footnote{The average gains for \textit{Loot} considers only its first 50 frames. For some frames of \textit{Man}, the SR inserted geometric artifacts orthogonal to the PC normals, perfectly recovering the geometry in a point-to-plane sense \protect{\cite{article:gpsnr_vetro:icip2017}}. This resulted in infinite GPSNR.}
For all sequences, our method achieves a superior performance in inferring the high-frequency. 
The \textit{Man} sequence benefited the most from the inferred high-frequency, while \textit{Phil} achieved the most modest enhancement.
The observed trend is that the more complex\footnote{In terms of bits per voxel to encode the full octree~\cite{article:garcia_queiroz:icip2017}.} the geometry, the greater potential to infer the high-frequency. 

\begin{figure}[h]
\centering
\begin{minipage}[h]{0.83\columnwidth}
\centering
\centerline{\includegraphics[width=\columnwidth]{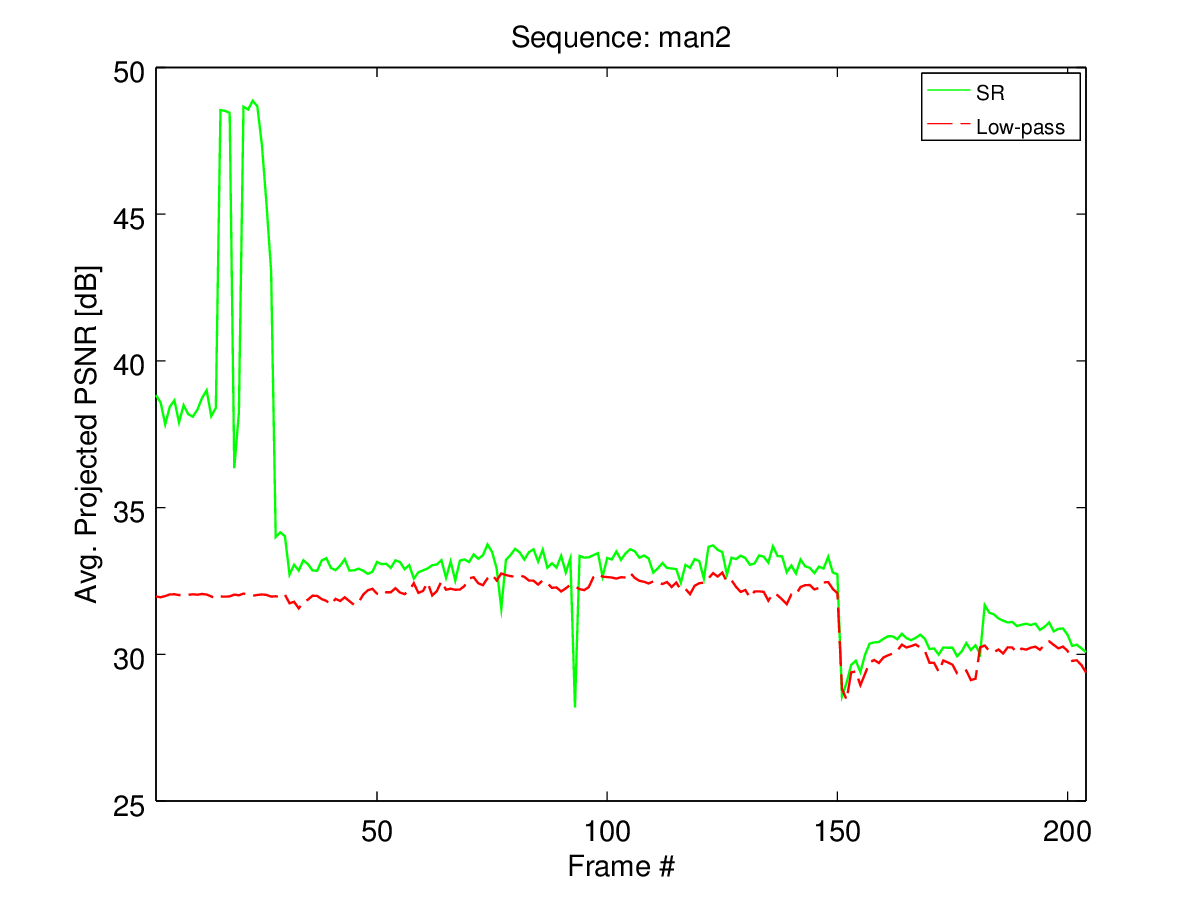}}
\centerline{(a)} 
\centerline{\includegraphics[width=\columnwidth]{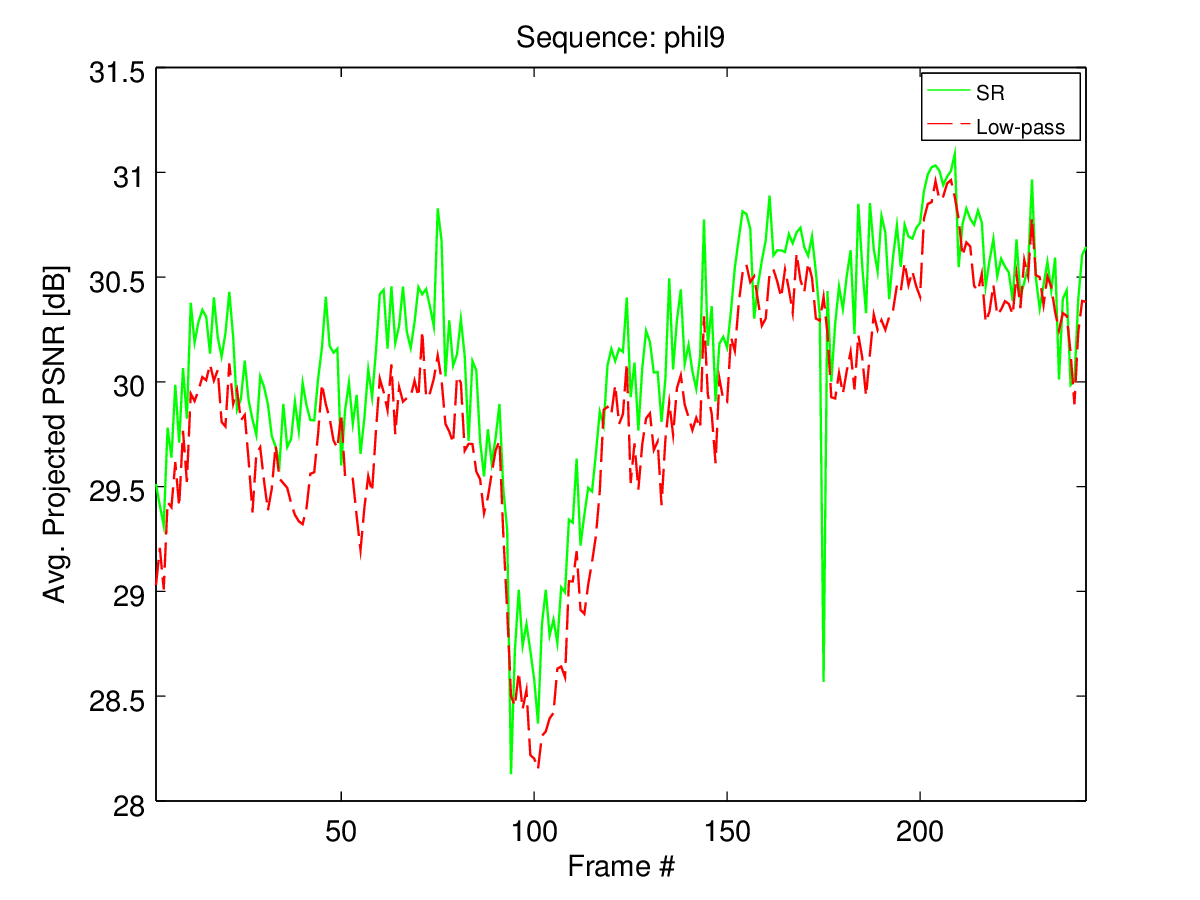}}
\centerline{(b)} 
\end{minipage}
\vspace{-2 ex}
\caption{\small Average PPSNR on a frame basis for SR and for the upsampled version (low-pass). Sequences: (a) \textit{Man}; (b) \textit{Phil}.}
\label{fig:results_man_phil}
\end{figure}

Figures~\ref{fig:results_man_phil}(a) and (b) present the PPSNR on a frame-by-frame basis for the low-pass point-cloud version and for the SR version for sequences \textit{Man} and \textit{Phil}, respectively. Figure~\ref{fig:results_man_phil}(a) shows the best high-frequency inference observed in the test set.
This is due to a relative lack of motion in sequence \textit{Man} on its first 30 frames; on the other side, an abrupt scene change around the 150th frame penalizes the quality of both versions (low-pass and SR). 
Despite the more challenging scenario in Fig.~\ref{fig:results_man_phil}(b), the SR framework yields better enhancement than the low-pass signal, on average.
\vspace{-2 ex}

\vspace{-2 ex}
\begin{figure*}[ht]
\begin{minipage}[t]{2.06\columnwidth}
\centering
\centerline{\includegraphics[width=\columnwidth]{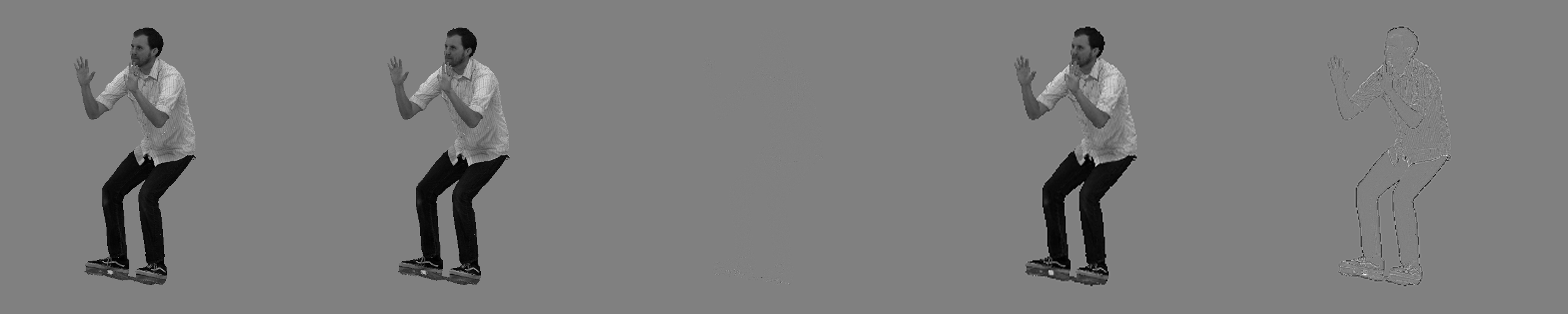}}
\centerline{(a)} 
\centerline{\includegraphics[width=\columnwidth]{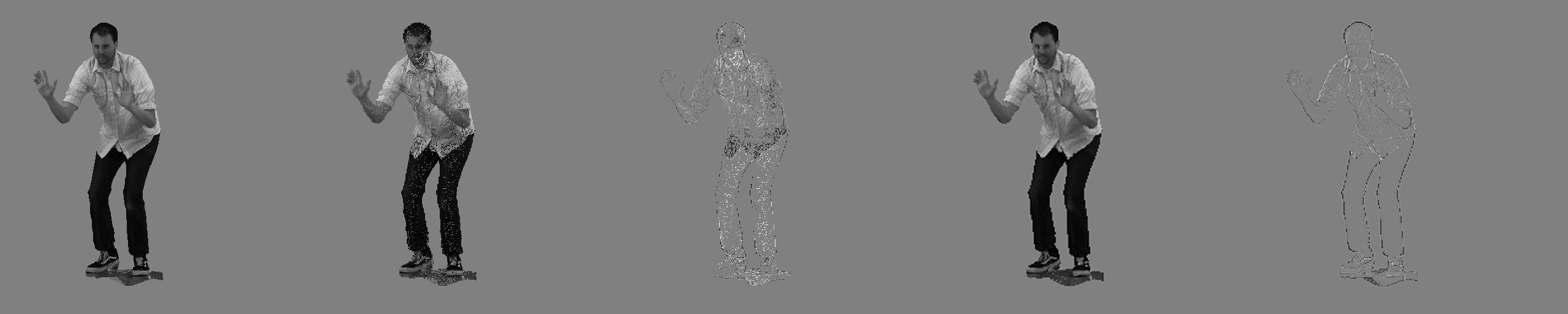}}
\centerline{(b)} 
\centerline{\includegraphics[width=\columnwidth]{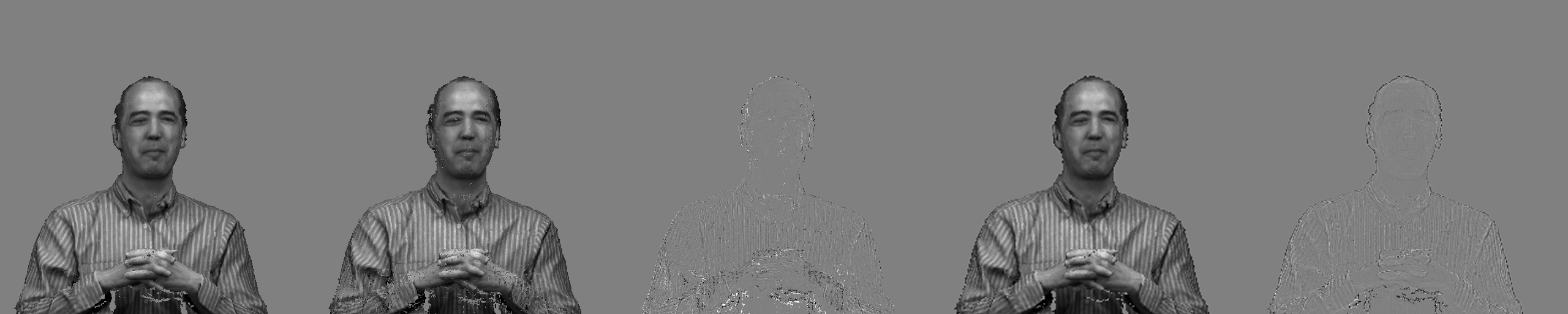}}
\centerline{(c)} 
\end{minipage}
\caption{\small Point-cloud projections for sequences (a)-(b) \textit{Man}, frames 23 and 93, and (c) \textit{Phil}, frame 175. For each image, from left to right, the columns correspond to the projections of: the original signal, the super-resolved signal, the residue of the super-resolved signal, the low-pass signal and the residue of the low-pass signal.}
\label{fig:projections_man_phil}
\end{figure*}

Figures~\ref{fig:projections_man_phil}(a)-(c) allow for the subjective evaluation of some point-cloud projections for sequences \textit{Man} and \textit{Phil}. 
Figure~\ref{fig:projections_man_phil}(a) shows the best SR performance for the test set, with an average 16.9 dB PPSNR gain over the low-pass version, mainly due to low movement of the test subject. The worst performance can be seen in Figs.~\ref{fig:projections_man_phil}(b) and (c), with average PPSNR losses of 4.12 and 1.84 dB over their low-pass versions, respectively.

\begin{table}[h]
\vspace{-4 ex}
\centering
\small
\caption{\small SR performance results. PPSNR-SR and PPSNR-LP stand for the average projected PSNR of the SR signal and the low-pass version, respectively. All values are in dB.}
\begin{tabular}{ c | c | c }
\textbf{Sequence} & PPSNR-SR & PPSNR-LP  \\ \hline
\textit{Andrew} & 30.83 & 30.17 \\ \hline
\textit{David} & 30.91 & 29.90 \\ \hline
\textit{Loot} & 41.61 & 39.73 \\ \hline
\textit{Man} & 33.52 & 31.59 \\ \hline
\textit{Phil} & 30.15 & 29.89 \\ \hline
\textit{Ricardo} & 33.60 & 32.36\\ \hline
\textit{Sarah} & 31.90 & 30.72
\end{tabular}
\label{tab:res1}
\end{table}

\begin{table}[h]
\vspace{-4 ex}
\centering
\small
\caption{\small SR performance improvements. PPSNR Gains and GPSNR Gains stand for the average gains in projected PSNR and in geometric quality metric~\cite{article:gpsnr_vetro:icip2017, article:queiroz_chou:tip2017}, respectively. All values are in dB.
	}
\begin{tabular}{ c | c | c }
\textbf{Sequence} & \textbf{PPSNR Gains} & \textbf{GPSNR Gains} \\ \hline
\textit{Andrew} & 0.76 & 4.99  \\ \hline
\textit{David} & 1.01  & 4.25\\ \hline
\textit{Loot} & 1.84  & 5.40\\ \hline
\textit{Man} & 1.93 &  $\infty$\\ \hline
\textit{Phil} & 0.27  & 4.61 \\ \hline
\textit{Ricardo} & 1.24 & 5.16 \\ \hline
\textit{Sarah} & 1.18 & 4.64 \\ \hline
\textit{Average} & \textbf{1.18} & \textbf{4.84}
\end{tabular}
\label{tab:res2}
\end{table}

\vspace{-2 ex}

\vspace{-3 ex}

\section{Conclusions}
\label{sec:conc}
\small
In this paper, an example based super-resolution framework to infer the high-frequency content of a voxelized point-cloud was presented.
Based on an already efficient point-cloud representation~\cite{octree}, we benefited from its inherent scalability in resolution to explore similarities between point-cloud frames of test sequences.
Experiments carried with seven point-cloud sequences show that the proposed method is able to successfully infer the high-frequency content for all the test sequences, yielding an average improvement of 1.18 dB when compared to a low-pass version of the test sequences.
These results can benefit a point-cloud encoding framework, for efficient transmission, error concealment or even storage.



\end{document}